\documentclass[aps, prl, showpacs,amsmath,amssymb, preprint, floatfix, superscriptaddress]{revtex4}
\usepackage{graphicx}
\usepackage{bm}
\begin{document}

\title{Structure change, layer sliding, and metallization in high-pressure MoS$_{2}$}

\author{Liliana Hromadov\'{a}}
\affiliation{Department of Experimental Physics, Comenius University, 
Mlynsk\'{a} Dolina F2, 842 48 Bratislava, Slovakia}
\affiliation{International School for Advanced Studies (SISSA) and CNR-IOM Democritos, Via Bonomea 265, I-34136 Trieste, Italy}

\author{Roman Marto\v{n}\'{a}k}
\email{martonak@fmph.uniba.sk}
\affiliation{Department of Experimental Physics, Comenius University, 
Mlynsk\'{a} Dolina F2, 842 48 Bratislava, Slovakia}

\author{Erio Tosatti}
\affiliation{International School for Advanced Studies (SISSA) and CNR-IOM Democritos, Via Bonomea 265, I-34136 Trieste, Italy}
\affiliation{The Abdus Salam International Centre for Theoretical Physics
  (ICTP), Strada Costiera 11, I-34151 Trieste, Italy}


\date{\today}

\begin{abstract}
  Based on ab initio calculations and metadynamics simulations, we predict
  that 2H-MoS$_2$, a layered insulator, will metallize under pressures in
  excess of 20-30 GPa. In the same pressure range, simulations and enthalpy
  optimization predict a structural transition. Free mutual sliding of
  layers takes place at this transition, the original 2H$_c$ stacking
  changing to a 2H$_a$ stacking typical of 2H-NbSe$_2$, an event explaining
  for the first time previously mysterious X-ray diffraction and Raman
  spectroscopy data. Phonon and electron phonon calculations suggest that
  pristine MoS$_2$, once metallized, will require ultrahigh pressures in
  order to develop superconductivity.
\end{abstract}

\maketitle

Transition metal dichalcogenides MX$_2$, where M is a transition metal and
X=S, Se and Te, are layered materials displaying a variety of electronic
behavior from insulating to charge-density-wave to metallic to
superconducting, with a rich scenario of phase transitions as a function of
external parameters. Intercalation of electron-donating atomic species
between the weakly bonded MX$_2$ layers is facile, turning the insulators
to metals which are interesting catalysts, and which also often
superconduct at cryogenic temperatures.  Perhaps the best known case in
point is that of MoS$_{2}$, a semiconductor with an indirect electronic
band gap of 1.29 eV \cite{gmelin}, which in its pristine state is well
known as a lubricant \cite{lubricant}, and more recently as a potential
photovoltaic \cite{Heinz} and single-layer transistor
\cite{layer_transistor} material. MoS$_{2}$ metallizes under intercalation
of alkalis and alkaline earths \cite{somoano}, achieving superconducting
temperatures up to 6 K. It has very recently also been shown to
superconduct at even higher temperatures upon electric double layer (EDL)
field doping \cite{Taniguchi2012,Ye}.

High pressure provides, besides doping, another general route to transform
insulators to metals. Recently it was shown that bilayer sheet of MoS$_{2}$
undergoes semiconductor-metal transition upon vertical compressive
pressure\cite{Bhattacharyya2012}. The possibility that MoS$_{2}$ might
metallize under pressure is also suggested by a negative pressure
coefficient of resistivity~\cite{resistivity} indicating gap shrinking,
$dE_G/dP < 0$.  Whether and how high pressure gap closing and metallization
in bulk could be achieved is, however, an open question, also because X-ray
diffraction \cite{aksoy} and Raman spectroscopy \cite{Livneh2010} data
indicate a structural transition taking place between 20 and 30 GPa to
another phase of unknown structure and properties. What is the structural
transition, whether high pressure metallization and possibly
superconductivity will eventually occur or not, and what could be the
interplay among these structural and electronic phenomena in an initially
insulating material, are all open questions of considerable interest, for
which MoS$_{2}$ can play a prototypical role.  We alternated and combined
well established first principles density functional theory (DFT)
calculations and ab initio metadynamics (AIMtD) simulations
\cite{PhysRevLett.90.075503,2006NatMa} to predict and clarify the
simultaneous evolution of structural and electronic properties of pristine
MoS$_{2}$ under pressure.

AIMtD simulation -- where the supercell parameters act as collective
variables -- is a powerful computational tool to discover new and competing
solid phases \cite{csp_wiley2010,epjb2011}. For ab initio molecular
dynamics and structural relaxations we employed VASP (Vienna ab initio
simulation package) \cite{vasp} with standard scalar relativistic PAW
pseudopotentials (no spin-orbit included)\cite{PhysRevB.59.1758} and cutoff
of 340 eV. MD simulations were performed using a $72$-atoms $2 \sqrt(3)
\times 2 \sqrt(3)$ supercell and 2$\times$2$\times$2
Monkhorst-Pack\cite{mpgrid} (MP) k-point sampling grid. For structural
relaxations we used the 6-atoms unit cell and a $9\times9\times5$ MP grid.
The electronic structure calculations and structural optimizations were
independently conducted using a variety of exchange-correlation
functionals: LDA, GGA with PBE parametrization \cite{PBE}, hybrid
functionals B3LYP \cite{Becke,Stephens} and HSE06 \cite{krukau}. Van der
Waals interactions were included in the Grimme approximation \cite{Grimme}
in the initial calculations at zero pressure, where they made an important
contribution to the interplanar attraction, but were subsequently removed
at high pressures. It turned out that at pressures above 5 GPa, in a regime
where interlayer interactions are repulsive it is GGA with PBE potential
and without van der Waals corrections that provides lattice parameters
closest to experimental data.

Within that approximation we thus carried out accurate total energy-based
structural relaxations at increasing pressures, refining the properties of
the phase(s) discovered with AIMtD dynamical simulations. That combined
approach produced the results which we now describe.

%%%%%%%%%%%%%%%%%%%%%%%%%%%%%%%%%%%%%%%%%%%%%%%%%%%%%%%%%%%%%%%%%%%%%%%%%%

\begin{figure}[htpb]
  \includegraphics*[width=8.8cm]{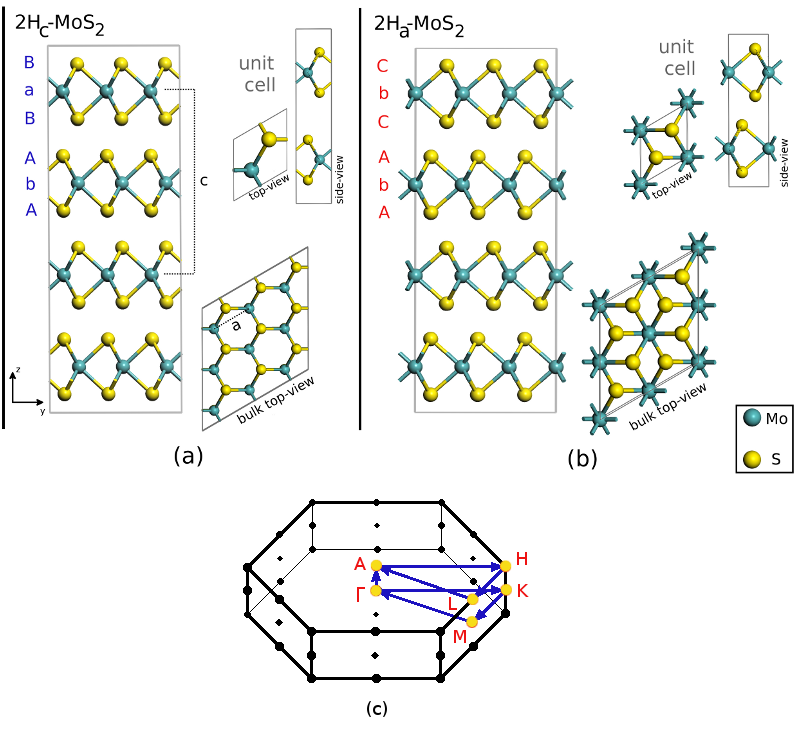}
  \caption{Structures of 2H$_c$-MoS$_2$ (a) and 2H$_a$-MoS$_2$ (b).
    I.~Brillouin zone for both structures\cite{xcrysden} (c).}
 \label{fig:structures}
\end{figure}

%%%%%%%%%%%%%%%%%%%%%%%%%%%%%%%%%%%%%%%%%%%%%%%%%%%%%%%%%%%%%%%%%%%%%%%%%%%

\begin{figure}[htpb]
  \includegraphics*[width=8.4cm]{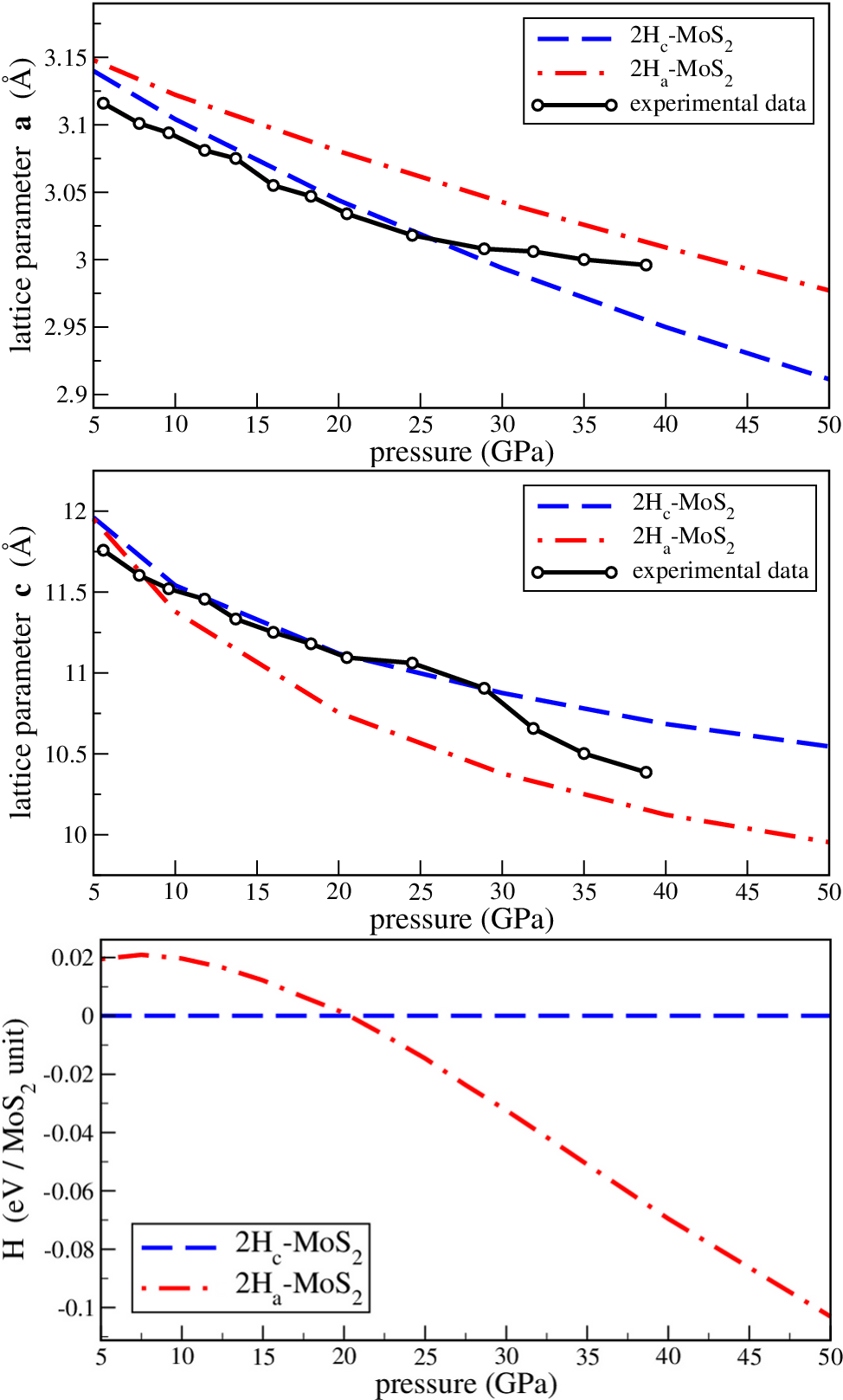}
  \caption{Calculated lattice parameters $a$ (up) and $c$ (middle), and
    relative enthalpies (bottom) of 2H$_a$- and 2H$_c$-MoS$_2$ structures
    as a function of pressure. The experimental data taken from Table 1 in
    Ref.\cite{aksoy} are in good agreement with calculated crossing of
    enthalpies at $p \sim 20$ GPa as shown in the bottom picture.}
 \label{fig:par_enth}
\end{figure}

\begin{figure}[htpb]
  \includegraphics*[width=8.6cm]{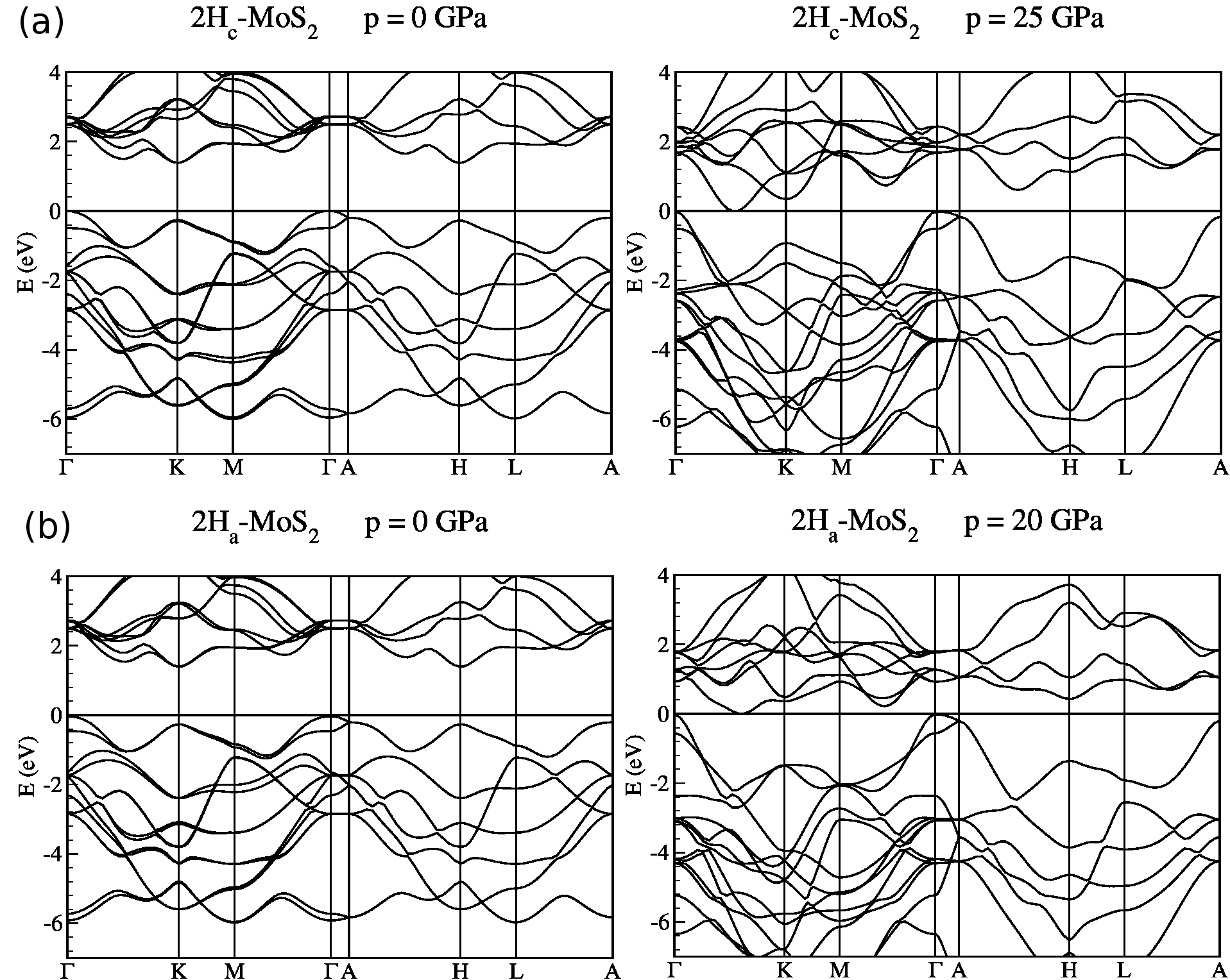}
  \caption{Calculated band structures of 2H$_c$- (a) and 2H$_a$-MoS$_2$ (b)
    at selected pressures.}
 \label{fig:bands}
\end{figure}

Static relaxation of the initial structure 2H$_c$-MoS$_2$ (space group
$P6_{3}/mmc$) of Fig.\ref{fig:structures} (a) \footnote{We adopt here the
  notation of Toledano et al.\cite{toledano}, while in earlier works such
  as, e.g., Ref.\cite{wickman}, this structure was denoted as 2H$_b$.} upon
increasing pressure produced no instability towards other phases up to 50
GPa.  The unit cell dimensions shrank anisotropically
(Fig.\ref{fig:par_enth}) as expected from a softer c-axis and harder a,b
axes. The indirect electronic band gap also shrank as expected, and
eventually closed near 25 GPa with an electron-hole wavevector $\delta k
\approx \frac{1}{2}K$ (Fig.\ref{fig:bands}) where $K$ =
($\frac{1}{3}$,$\frac{1}{3}$,0) is a Brillouin Zone edge point
(Fig.\ref{fig:structures}(c)) thus predicting band overlap metallization of
2H$_c$-MoS$_2$ above 25 GPa. This result agrees with the very recent finding in
Ref.\cite{Guo2012}.  Alternative DFT calculations and relaxations carried
out with the HSE06 \cite{krukau} and the B3LYP \cite{Becke, Stephens}
approximations yielded a metallization pressure of 35 and 55 GPa,
respectively. Since these approaches, especially the latter, are known to
underestimate metallicity, whereas GGA/PBE may overestimate it, we conclude
that pressure induced metallization of 2H$_c$-MoS$_2$ should be placed
between 25 and 35 GPa.

Extending in this pressure region our calculations to the $72$-atoms $2
\sqrt(3) \times 2 \sqrt(3)$ supercell in place of the initial $6$-atom unit
cell, we further explored the possible onset of a periodic lattice
distortion or charge-density-wave phase (more properly called an "excitonic
insulator" \cite{PhysRev.158.462}). We performed structural optimizations
of samples with small random displacements of atoms and looked for the
possible onset of the accompanying periodic lattice distortion, but failed
to find any. In fact, the possible occurrence of an excitonic insulator
driven superstructure in a narrow pressure interval near the
insulator-metal transition, an occurrence which can be considered quite
probable, cannot seriously be established within our methods, and more
refined treatments of exchange will be called for in the future.

%%%%%%%%%%%%%%%%%%%%%%%%%%%%%%%%%%%%%%%%%%%%%%%%%%%%%%%%%%%%%%%%%%%%%%%%%%%

%%%%%%%%%%%%%%%%%%%%%%%%%%%%%%%%%%%%%%%%%%%%%%%%%%%%%%%%%%%%%%%%%%%%%%%%%%

%%%%%%%%%%%%%%%%%%%%%%%%%%%%%%%%%%%%%%%%%%%%%%%%%%%%%%%%%%%%%%%%%%%%%%%%%%% 

\begin{figure}[htpb]
  \includegraphics*[width=8.6 cm]{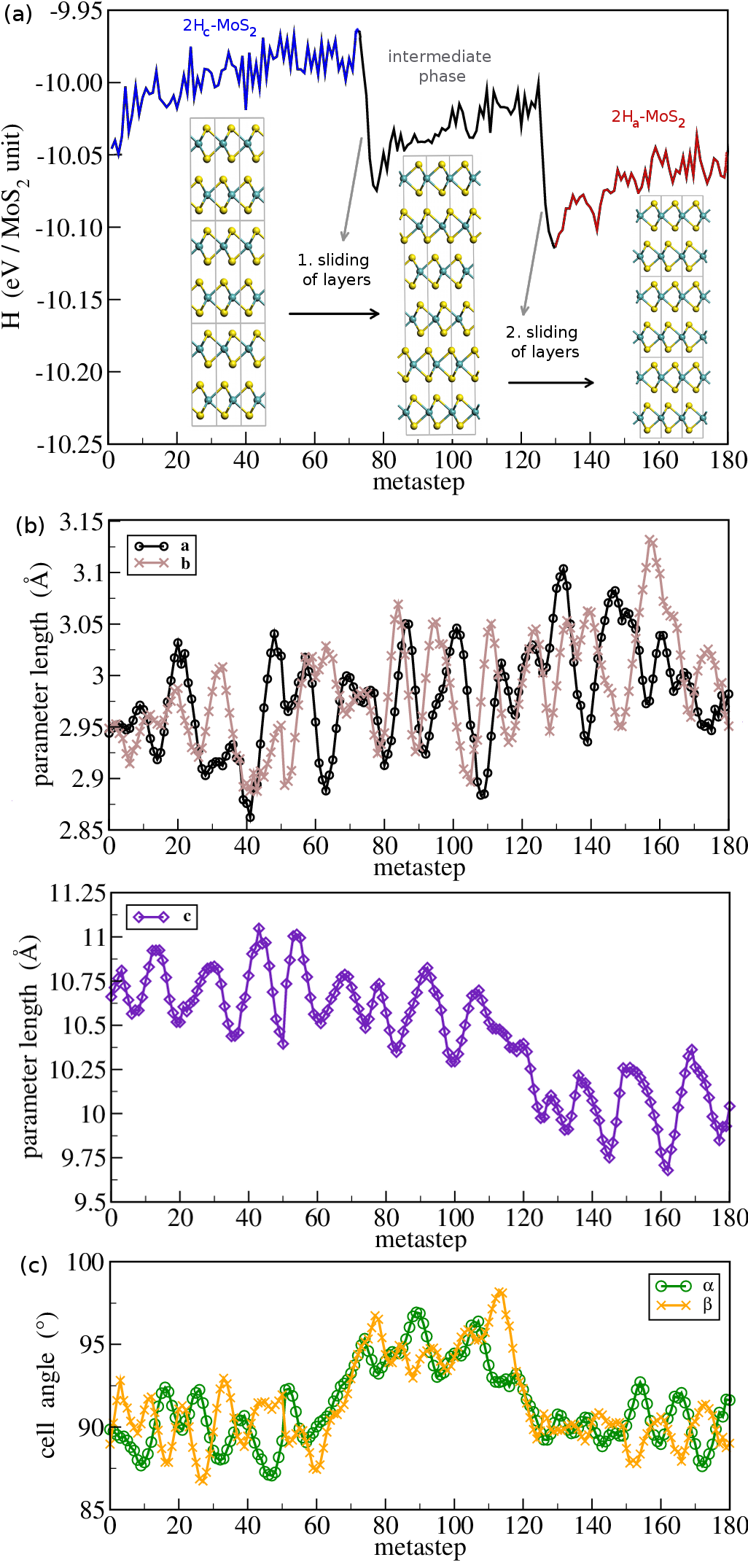}
  \caption{(a) Evolution of enthalpy in the metadynamics simulation during
    transformation from 2H$_c$-MoS$_2$ to 2H$_a$-MoS$_2$ via an
    intermediate structure. (b) Evolution of unit cell parameters $a, b,
    c$. The parameters $a, b$ are equal to the size of the 72-atoms
    supercell in respective directions divided by $2 \sqrt{3}$. (c)
    Evolution of supercell angles $\alpha, \beta$. The angle $\gamma$ (not
    shown) fluctuates around 120$^{\circ}$ and does not undergo a
    significant change.}
 \label{fig:metadynamics}
\end{figure}

We focused so far on the pressure evolution of the 2H$_c$-MoS$_2$ initial
structure.  However, the local stability which we found upon relaxation is
no guarantee for global stability at higher pressures.  The AIMtD
simulations, where the supercell vectors are biased away from the initial
structure towards newer ones, can reveal the possible structures missed by
straight relaxation.  When applied to 2H$_c$-MoS$_2$ in the 72-atom
supercell at $p=40$ GPa, it led to two successive and surprising sliding
events shown in Fig.\ref{fig:metadynamics}. In the first event after
metastep 73 the two layers within the supercell shifted with respect to
each other so that the Mo atom $(x,y)$ coordinates in the two layers became
coincident (Fig.\ref{fig:metadynamics}(a)).  At the same time the supercell
developed a tilt: the $\alpha$ and $\beta$ angles moved away from right
angle as can be seen in Fig.\ref{fig:metadynamics}(c). In this intermediate
structure the parameter $a$ increased and the parameter $c$ decreased from
their original values (Fig.\ref{fig:metadynamics}(b)). In the second event
following metastep 125 the $\alpha$ and $\beta$ angles returned to the
original 90 degrees while the parameters $a$ and $c$ further increased and
decreased, respectively.  The final structure was relaxed at $p=40$ GPa and
T=0 and was identified as 2H$_a$ -- another 2H polytype with the same space
group, but where the layer stacking is now AbA CbC in place of the initial
AbA BaB stacking of 2H$_c$\cite{toledano,wickman}. In the new structure --
typical of e.g., 2H-NbSe$_2$ -- all Mo atoms share the same $(x,y)$
coordinates.  Thus the pressure induced easy sliding of layers -- a
"superlubric" event, possibly related to MoS$_2$'s lubricant properties --
transformed the 2H$_c$ structure of MoS$_2$ to 2H$_a$.

%%%%%%%%%%%%%%%%%%%%%%%%%%%%%%%%%%%%%%%%%%%%%%%%%%%%%%%%%%%%%%%%%%%%%%%%%%

%%%%%%%%%%%%%%%%%%%%%%%%%%%%%%%%%%%%%%%%%%%%%%%%%%%%%%%%%%%%%%%%%%%%%%%%%%%

Through accurate structural relaxations and electronic structure
calculations we refined the cell parameters, the enthalpy and the
electronic band structure of 2H$_a$-MoS$_2$, with results shown in
Figs.~\ref{fig:par_enth}, \ref{fig:bands}. The 2H$_c$ and 2H$_a$ enthalpies
cross near 20 GPa, justifying the structural transition discovered by
AIMtD.\footnote{Zero-point motion and entropic contributions for both
  phases were estimated within the quasiharmonic approximation at $p=10$
  GPa and $T=300$ K. The difference between the two phases was smaller than
  1 meV/atom and therefore these contributions were neglected in the
  figure.}  The new 2H$_a$-MoS$_{2}$ structure has at $p=20$ GPa lattice
parameters $a= 3.080 \; \AA, c = 10.754 \; \AA$ with Mo atoms at Wyckoff
position 2(b) and S atoms at 4(f) with $z = 0.8947$.  As the band
structures of Fig.\ref{fig:bands} (b) show, the new phase 2H$_a$-MoS$_{2}$,
also semiconducting at low pressures, undergoes a pressure-induced band
overlap metallization now near 20 GPa. Structurally, 2H$_a$ is less
anisotropic than the initial 2H$_c$ structure, the new interlayer spacing
(reflected in the c axis length) smaller, the in-plane interatomic spacing
(reflected by $a, b$ lengths) larger.

\begin{figure}[htpb]
  \includegraphics*[width=8.6cm]{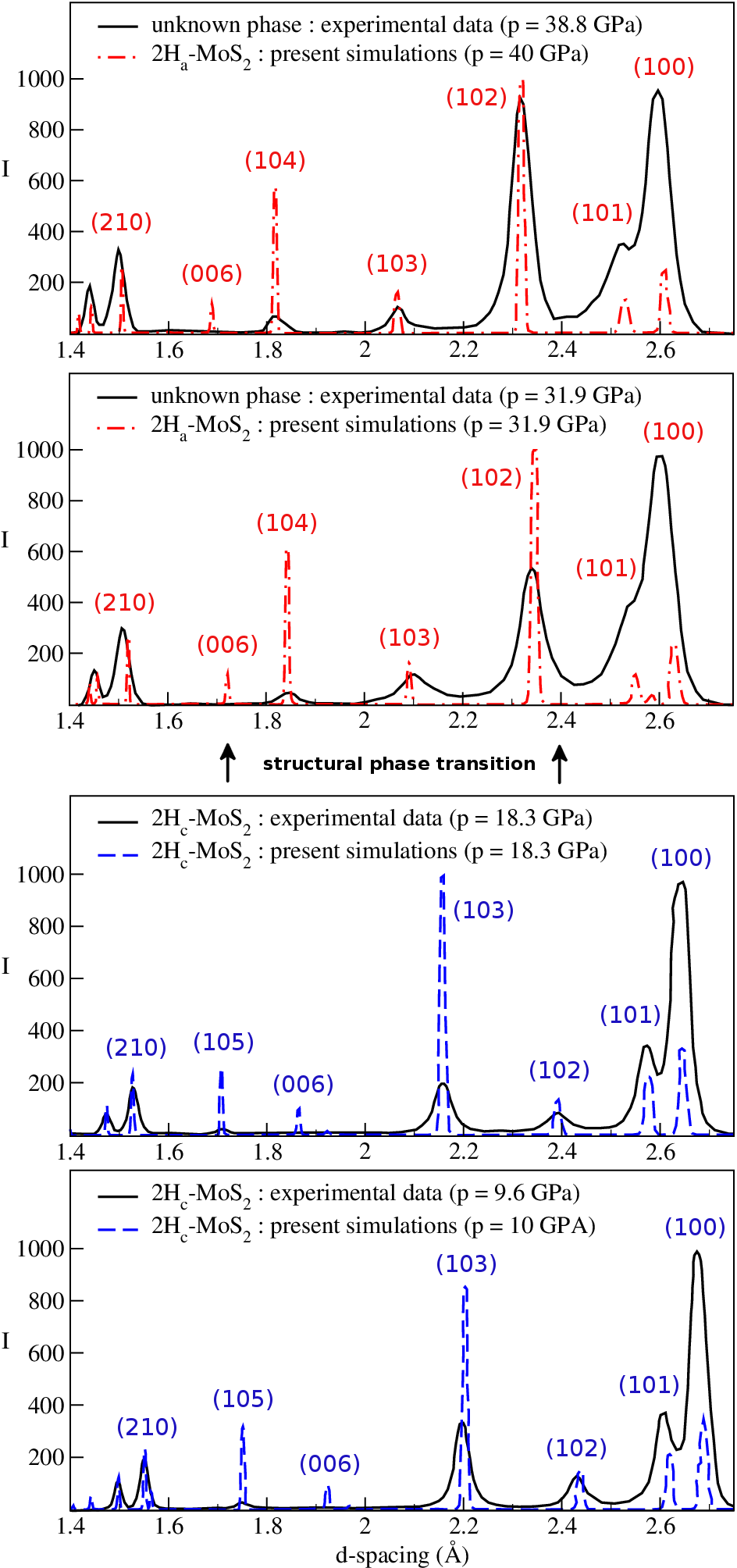}
  \caption{Comparison of experimental \cite{aksoy} and calculated X-ray
    diffraction patterns for 2H$_c$ structure below 20 GPa and 2H$_a$
    structure above 20 GPa.}
 \label{fig:diffraction}
\end{figure}

At this point comes an important contact with experiment.  Few years ago
Aksoy et al.~\cite{aksoy} had reported an unspecified structural
transformation taking place in 2H$_c$-MoS$_2$ between 20 and 30 GPa,
without change of space group symmetry. When we overlap their measured
lattice parameters $a$ and $c$ with our calculated ones in
Fig.\ref{fig:par_enth} the conclusion that their observed transformation is
precisely the 2H$_c$ to 2H$_a$ transition strongly suggests itself. To
further nail down that conclusion we calculated X-ray powder patterns for
2H$_c$-MoS$_2$ and 2H$_a$-MoS$_2$ and compared them in
Fig.\ref{fig:diffraction} with the measured ones. Again, the agreement is
quite good. The pressure induced onset of the 2H$_a$ phase is heralded by
the birth of a (104) reflection (indexed as "006" in Ref. \cite{aksoy}) and
a much stronger (102) reflection.  The simultaneous demise of the
2H$_c$-MoS$_2$ phase is signaled by the disappearance of (105) and the drop
of (103) reflections.

%%%%%%%%%%%%%%%%%%%%%%%%%%%%%%%%%%%%%%%%%%%%%%%%%%%%%%%%%%%%%%%%%%%%%%%%%%

%%%%%%%%%%%%%%%%%%%%%%%%%%%%%%%%%%%%%%%%%%%%%%%%%%%%%%%%%%%%%%%%%%%%%%%%%

\begin{figure}[htpb]
  \includegraphics*[width=8.6cm]{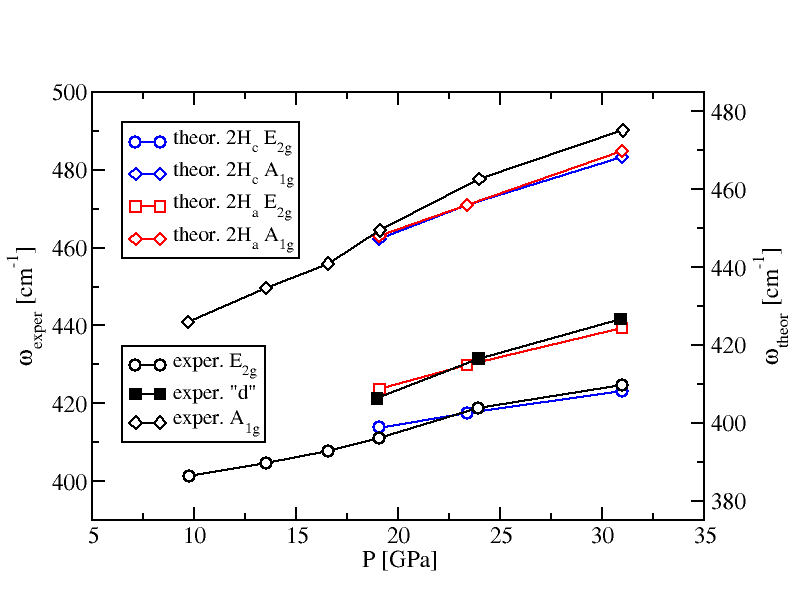}
  \caption{Comparison of experimental\cite{Livneh2010} and calculated peak
    positions of Raman spectra for 2H$_c$ and 2H$_a$ structures at
    different pressures. The vertical axis for theoretical data (right) has
    been shifted with respect to the axis for experimental data (left) by
    15 cm$^{-1}$ upwards to account for a small constant difference between
    the two datasets.}
 \label{fig:Raman}
\end{figure}

Raman spectroscopy data \cite{Livneh2010} are also available for MoS$_2$ in
the pressure interval up to 31 GPa. At $p=19.1$ GPa a new peak (called
``d'' band, Figs.~5, 6 in Ref.\cite{Livneh2010}) appears, first as a
shoulder on the high-frequency side of the original E$_{2g}$ peak.  As
pressure grows its weight increases at the expense of the original E$_{2g}$
peak and at 23.4 and 31 GPa both peaks are clearly visible in the spectra.
Using the Quantum Espresso code\cite{quantum_espresso} we calculated the
frequencies of Raman active E$_{2g}$ and A$_{1g}$ phonon modes of
2H$_c$-MoS$_2$ and 2H$_a$-MoS$_2$ at the same experimental pressures of
19.1, 23.4 and 31 GPa.\footnote{We used the PBE PAW pseudopotentials
  Mo.pbe-spn-kjpaw.UPF and S.pbe-n-kjpaw.UPF.}  Comparison with
experimental data\cite{Livneh2010} in Fig.\ref{fig:Raman} suggests that the
``d'' band coincides with the E$_{2g}$ mode of the emergent 2H$_a$ phase.
On the other hand the difference of frequency of the higher A$_{1g}$ mode
in both phases is very small which explains why this peak is not observed
to split in the experiment. Apart from a small systematic difference of
about 15 cm$^{-1}$ between experimental and calculated peak positions the
overall agreement is very good. This provides further strong support for
the 2H$_c$ $\rightarrow$ 2H$_a$ structural transition starting at 20 GPa
with both phases coexisting up to the highest pressure of 31 GPa.

%%%%%%%%%%%%%%%%%%%%%%%%%%%%%%%%%%%%%%%%%%%%%%%%%%%%%%%%%%%%%%%%%%%%%%%%%

%%%%%%%%%%%%%%%%%%%%%%%%%%%%%%%%%%%%%%%%%%%%%%%%%%%%%%%%%%%%%%%%%%%%%%%%%

We can now stretch beyond experimental observations, which stop just below
40 GPa, by extending simulations to higher pressures.\footnote{Above 50 GPa
  we used the hard version of VASP PAW pseudopotentials.} At 120 GPa,
searching for newer potential structural transitions, we carried out
another metadynamics simulation but the 2H$_a$ structure within the 72-atom
supercell remained stable and did not show any transition. We further
performed a structural relaxation of the 2H$_a$ structure within the 6-atom
unit cell at ultrahigh pressure of 200 GPa and again failed to find any new
phase, indicating good local stability of 2H$_a$.

\begin{figure}[htpb]
  \includegraphics*[width=8.6cm]{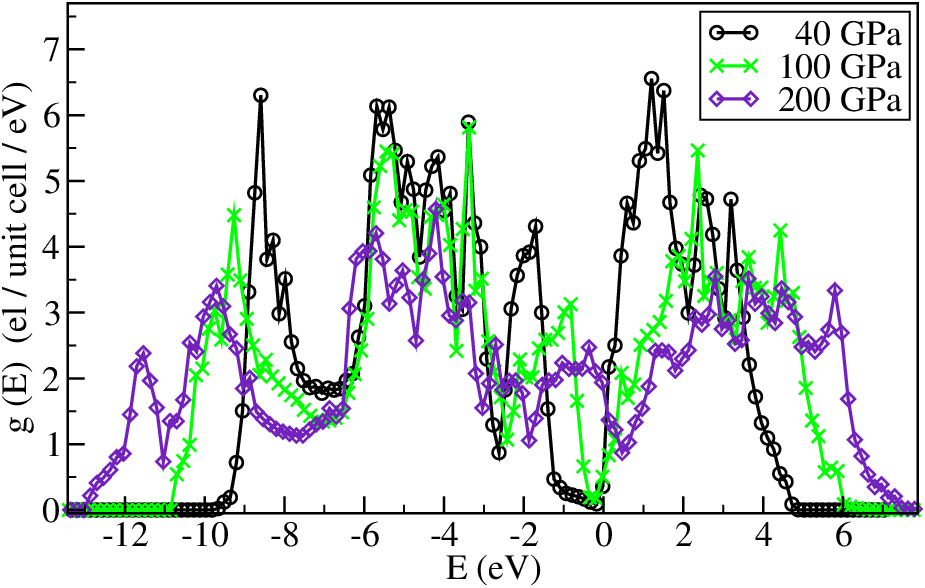}
  \caption{2H$_a$-MoS$_2$ electronic density of states at selected
    pressures.}
 \label{fig:edos}
\end{figure}

Having discovered that MoS$_2$ turns metallic above 20-30 GPa, it is a
relevant question whether it will superconduct. Inspection of the
electronic density of states (DOS) at the Fermi level $n(E_F)$
(Fig.~\ref{fig:edos}) shows that the metallicity of 2H$_a$-MoS$_2$,
marginal near 20 GPa, increases only very mildly with pressure.  Poor
metallicity is probably the reason why Raman spectra are still well defined
even at 31 GPa. Using again the Quantum Espresso
code\cite{quantum_espresso} we calculated besides the DOS, also the
dimensionless electron-phonon coupling strength $\lambda$ and the effective
logarithmic average phonon frequency $\omega_{log}$ at several pressures.
As Table \ref{table:superconducting} shows, superconducting temperatures
estimated by the Allen-Dynes formula \cite{PhysRevB.12.905} are not
encouraging. Even for optimistic values for the Coulomb pseudopotential
$\mu*$, superconductivity is predicted to appear only at ultrahigh
pressures well beyond 100 GPa.

%%%%%%%%%%%%%%%%%%%%%%%%%%%%%%%%%%%%%%%%%%%%%%%%%%%%%%%%%%%%%%%%%%

%%%%%%%%%%%%%%%%%%%%%%%%%%%%%%%%%%%%%%%%%%%%%%%%%%%%%%%%%%%%%%%%%%%%%%

\begin{table}[h]
  \caption{Values of $p$ [GPa], $\lambda$, $\omega_{log}$ [K], $n(E_F)$ [electrons/unit
    cell/eV], $T_c(\mu^{*}=0.1)$ [K], $T_c(\mu^{*}=0.05)$ [K] for 2H$_c$-MoS$_2$ and 2H$_a$-MoS$_2$.}
  \centering
                \begin{tabular}{|l |l| l| l| l| l| l|}
                        \hline \hline
                phase & $p$  & $\lambda$ & $\omega_{log}$ & $n(E_F)$ &
                $T_c(0.1)$ & $T_c(0.05)$ \\ [0.5ex]
                         \hline
                           2H$_c$ &   40 & 0.095 & 384 &  0.45 & 0.0 & 0.0 \\
                           2H$_c$ & 100 & 0.118 & 518 &  0.68 & 0.0 & 0.0 \\
                           2H$_a$ &   40 & 0.078 & 388 &  0.44 & 0.0 & 0.0 \\
                           2H$_a$ & 100 & 0.108 & 402 &  0.49 & 0.0 & 0.0 \\
                           2H$_a$ & 200 & 0.330 & 532 & 1.69 & 0.6 & 2.7 \\ [1ex]
                          \hline
                \end{tabular}
                \label{table:superconducting}
\end{table} 

%%%%%%%%%%%%%%%%%%%%%%%%%%%%%%%%%%%%%%%%%%%%%%%%%%%%%%%%%%%%%%%%%%%%%

Our overall physical conclusions are therefore that pressure will cause
MoS$_2$ layers to slide, causing a structural transition between $2H_c$ and
$2H_a$ polytypes, via a process akin to superlubric sliding.~\cite{vanossi}
In both structures, band-overlap metallization will take place at
relatively low pressures. After that, however, the broad cleft between
valence and conduction bands, both with large Mo $d$-band character, does
not shrink fast enough to produce a comparably large metallicity to that
generated by, e.g., alkali doping, or by EDL field doping
\cite{Taniguchi2012, Ye}.  High pressure MoS$_2$ is predicted to remain
semimetallic, and only prone to superconductivity well beyond 100 GPa.  The
contrast with the higher superconducting temperatures obtained by alkali
and by EDL electron doping stems first of all from the smaller DOS
achievable by pressure, but can also be related to a higher coupling of
electrons relative to holes, and to a dimensionality effect in the EDL
case.  One aspect that remains to be explored is the possible occurrence,
to be pursued with methods beyond those currently available, of excitonic
insulator driven charge or spin density waves in a narrow pressure range
close to the metallization pressure. If such a phase did exist, it is
likely to be preceded and/or followed by a superconducting region.

\begin{acknowledgments}
  E.T. thanks T. Kagayama, Y. Iwasa, A. Shukla, and M. Calandra for
  discussions and exchange of information. Work in Trieste was partly
  sponsored by EU-Japan Project LEMSUPER, and by Sinergia Contract
  CRSII2$_1$36287/1. L.H. and R.M were supported by the Slovak Research and
  Development Agency under Contract No.~APVV-0558-10 and by the project
  implementation 26220220004 within the Research $\&$ Development
  Operational Programme funded by the ERDF. Part of calculations were
  performed in the Computing Centre of the Slovak Academy of Sciences using
  the National Supercomputing Infrastructure supported from Structural
  Funds of EU. L.H. acknowledges a predoc fellowship held at SISSA during
  part of this project.
\end{acknowledgments}

\end{document}